\documentclass[final]{aipproc}

\layoutstyle{8x11double}

\newcommand\msun {M$_{\odot}$}
\def\approxgt{\ifmmode \rlap{$>$}{}_{{}_{{}_{\textstyle\sim}}} \else%
$\rlap{$>$}{}_{{}_{{}_{\textstyle\sim}}}$\fi} 
\def\approxlt{\ifmmode \rlap{$<$}{}_{{}_{{}_{\textstyle\sim}}} \else%
$\rlap{$<$}{}_{{}_{{}_{\textstyle\sim}}}$\fi}

\def\xmm{XMM-{\it Newton}}
\def\src{Cen~X--4}
\def\cxo{{\it Chandra}}

\begin{document}

\title{Constraining the neutron star equation of state using quiescent low--mass X--ray binaries}


\author{P.G. Jonker}{
  address={SRON, Netherlands Institute for Space Research, 3584~CA, Utrecht, The Netherlands}
  ,altaddress={Harvard--Smithsonian  Center for Astrophysics, Cambridge, MA~02138, Massachusetts, U.S.A.}
  ,altaddress={Astronomical Institute, Utrecht University, 3508 TA, Utrecht, The Netherlands}
}


%
%
%


\def\jnl@style{\it}
\def\aaref@jnl#1{{\jnl@style#1}}

\def\aaref@jnl#1{{\jnl@style#1}}

\def\aj{\aaref@jnl{AJ}}                   
\def\araa{\aaref@jnl{ARA\&A}}             
\def\apj{\aaref@jnl{ApJ}}                 
\def\apjl{\aaref@jnl{ApJ}}                
\def\apjs{\aaref@jnl{ApJS}}               
\def\ao{\aaref@jnl{Appl.~Opt.}}           
\def\apss{\aaref@jnl{Ap\&SS}}             
\def\aap{\aaref@jnl{A\&A}}                
\def\aapr{\aaref@jnl{A\&A~Rev.}}          
\def\aaps{\aaref@jnl{A\&AS}}              
\def\azh{\aaref@jnl{AZh}}                 
\def\baas{\aaref@jnl{BAAS}}               
\def\jrasc{\aaref@jnl{JRASC}}             
\def\memras{\aaref@jnl{MmRAS}}            
\def\mnras{\aaref@jnl{MNRAS}}             
\def\pra{\aaref@jnl{Phys.~Rev.~A}}        
\def\prb{\aaref@jnl{Phys.~Rev.~B}}        
\def\prc{\aaref@jnl{Phys.~Rev.~C}}        
\def\prd{\aaref@jnl{Phys.~Rev.~D}}        
\def\pre{\aaref@jnl{Phys.~Rev.~E}}        
\def\prl{\aaref@jnl{Phys.~Rev.~Lett.}}    
\def\pasp{\aaref@jnl{PASP}}               
\def\pasj{\aaref@jnl{PASJ}}               
\def\qjras{\aaref@jnl{QJRAS}}             
\def\skytel{\aaref@jnl{S\&T}}             
\def\solphys{\aaref@jnl{Sol.~Phys.}}      
\def\sovast{\aaref@jnl{Soviet~Ast.}}      
\def\ssr{\aaref@jnl{Space~Sci.~Rev.}}     
\def\zap{\aaref@jnl{ZAp}}                 
\def\nat{\aaref@jnl{Nature}}              
\def\iaucirc{\aaref@jnl{IAU~Circ.}}       
\def\aplett{\aaref@jnl{Astrophys.~Lett.}} 
\def\apspr{\aaref@jnl{Astrophys.~Space~Phys.~Res.}}
\def\bain{\aaref@jnl{Bull.~Astron.~Inst.~Netherlands}} 
\def\fcp{\aaref@jnl{Fund.~Cosmic~Phys.}}  
\def\gca{\aaref@jnl{Geochim.~Cosmochim.~Acta}}   
\def\grl{\aaref@jnl{Geophys.~Res.~Lett.}} 
\def\jcp{\aaref@jnl{J.~Chem.~Phys.}}      
\def\jgr{\aaref@jnl{J.~Geophys.~Res.}}    
\def\jqsrt{\aaref@jnl{J.~Quant.~Spec.~Radiat.~Transf.}}
\def\memsai{\aaref@jnl{Mem.~Soc.~Astron.~Italiana}}
\def\nphysa{\aaref@jnl{Nucl.~Phys.~A}}   
\def\physrep{\aaref@jnl{Phys.~Rep.}}   
\def\physscr{\aaref@jnl{Phys.~Scr}}   
\def\planss{\aaref@jnl{Planet.~Space~Sci.}}   
\def\procspie{\aaref@jnl{Proc.~SPIE}}   

\let\astap=\aap
\let\apjlett=\apjl
\let\apjsupp=\apjs
\let\applopt=\ao

\begin{abstract}

\cxo\, or \xmm\, observations of quiescent low--mass X--ray binaries can provide important constraints on the equation of state of neutron stars. The mass and radius of the neutron star can potentially be determined from fitting a neutron star atmosphere model to the observed X--ray spectrum. For a radius measurement it is of critical importance that the distance to the source is well constrained since the fractional uncertainty in the radius is at least as large as the fractional uncertainty in the distance. Uncertainties in modelling the neutron star atmosphere remain. At this stage it is not yet clear if the soft thermal component in the spectra of many quiescent X--ray binaries is variable on timescales too short to be accommodated by the cooling neutron star scenario. This can be tested with a long \xmm\, observation of the neutron star X-ray transient Cen~X--4 in quiescence. With such an observation one can use the Reflection Grating Spectrometer spectrum to constrain the interstellar extinction
to the source. This removes this parameter from the X-ray spectral fitting of
the EPIC pn and MOS spectra and allows one to investigate whether the variability
observed in the quiescent X--ray spectrum of this source is due to variations in
the soft thermal spectral component or variations in the power law spectral
component coupled with variations in ${\rm N_H}$. This will test whether the
soft thermal component can indeed be due to the hot thermal glow of the
neutron star. Irrespective of the outcome of such a study, the observed cooling in quiescence in sources for which the crust is significantly out of thermal equilibrium with the core due to a prolonged outburst, such as KS~1731--260, seem excellent candidates for mass and radius determinations through modelling the observed X--rays with a neutron star atmosphere model (the caveats about the source distance and atmosphere modelling do also apply here obviously and presently prevent one from obtaining such constraints). Finally, the fact that the soft thermal glow in sources such as SAX~J1808.4--3658 and 1H~1905+000 has not been detected in quiescence means that the neutron star cores of these sources must be cold. The most plausible explanation seems to be that the neutron stars are more massive than 1.4 \msun\, and cool via the direct URCA process.

\end{abstract}


\maketitle


\section{Introduction}

The majority of low--mass X--ray binaries is found to be transient -- the so called soft X--ray transients (here, I will just refer to them as X--ray transients or just transients, see e.g.~\cite{1997ApJ...491..312C}). The accretion rate in quiescence is much reduced or even stopped in these transient systems in quiescence. Before the \xmm\, and {\it Chandra} satellites were operational only a few neutron star X--ray transients had been studied in quiescence (e.g.~the systems Cen~X--4 and Aql~X--1; \cite{1987A&A...182...47V}, \cite{1995ApJ...442..358M}, \cite{1997A&A...324..941C}). Currently, many more systems have been studied in quiescence (see e.g.~\cite{2001ApJ...560L.159W}, \cite{2002ApJ...580..413R}, \cite{2001ApJ...560L.159W}, \cite{2002ApJ...575L..15C}, \cite{2003MNRAS.341..823J}, \cite{2004ApJ...610..933T}) and Cen~X--4 and Aql~X--1 have been studied in much more detail than was possible before (e.g.~\cite{2004ApJ...601..474C}, \cite{2002ApJ...577..346R}). As will become clear later, these observations can have a profound impact on an important area of astrophysics: determining the neutron star equation of state (EoS). This is one of the ultimate goals of the study of neutron stars.

Theoretically, one expects the neutron star to emit X--rays even after accretion has stopped (\cite{1998ApJ...504L..95B}). Due to the large heat capacity the temperature of the neutron star core is set on timescales of tens of thousands of years (\cite{2001ApJ...548L.175C}). The equilibrium core temperature depends on the balance between the heating and the cooling rate of the neutron star. The heating rate depends on the total amount of accreted baryons and the pycnonuclear reactions taking place a few hundred meters deep in the crust (\cite{1990A&A...227..431H}, \cite{1998ApJ...504L..95B}, \cite{2001ApJ...548L.175C}). The time--averaged mass accretion rate in neutron star transients can be derived from binary evolution models if the orbital period is known (e.g.~\cite{1962ApJ...136..312K}). The pycnonuclear reactions taking place in the neutron star crust are described in \cite{1969ApJ...155..183S} and \cite{2000ApJ...539..888K}. The cooling properties of the neutron star core depend on the equation of state (see e.g.~\cite{2004ARA&A..42..169Y},\cite{yakovthisvol}).

The resultant X--ray emission can be modelled by a neutron star atmosphere (NSA) model (\cite{1996A&A...315..141Z}, \cite{2002A&A...386.1001G}, \cite{2006ApJ...644.1090H}). The NSA models have four free parameters. The neutron star distance, mass, radius and temperature. Therefore, in theory, an NSA--fit provides means to measure the mass and radius of the neutron star and thus constrain the equation of state (EoS) of matter at supranuclear densities. In practice, spectra of neutron star transients in quiescence where the flux is
high enough to allow for a spectral study, are indeed well--fit by a neutron star
atmosphere model (NSA). Sometimes, additional emission is present at energies above
a few keV, often  quantified by a power--law. When systems are selected for which
the distance is well known (most notably sources in globular clusters), neutron star masses
and radii can be determined accurately (e.g.~\cite{2003ApJ...588..452H}, \cite{2007arXiv0708.3816W}). Indeed, values for the neutron star mass and radius consistent with those of canonical neutron stars are found, rendering support for the interpretation that the soft X-ray spectral component is due to the NSA. 

Besides fitting for the neutron star radius and mass in the NSA modelling the neutron star mass and possibly the neutron star equation of state can be constrained by investigating the neutron star temperature. The stringent upper limit on the NSA spectral component to the luminosity of SAX J1808.4--3658 ($\approxlt$10\%; \cite{2002ApJ...575L..15C}, \cite{2007ApJ...660.1424H}) and the deep limit on the luminosity and neutron star temperature in 1H~1905+000 (\cite{2006MNRAS.368.1803J}, \cite{2007ApJ...665L.147J}) hint at massive neutron stars in these two transients. The reasoning is as follows: the upper limits on the thermal spectral component imply low core temperatures, which in SAX J1808.4--3658 and 1H~1905+000 are so low that they imply a rapid release of the energy produced in the crust. This rapid release of energy can only occur via enhanced neutrino emission in neutron stars with masses larger than 1.6--1.7 $M_\odot$ (\cite{2004ARA&A..42..169Y}). Neutron stars with masses well above $1.4\,M_\odot$ cannot exist for so--called soft equations of state (EoS), in which matter at high densities is relatively compressible (e.g., due to a meson condensate or a transition between the hadron and quark--gluon phases). An important caveat is that the heating accretion history of the last several tens of thousands of years has to be determined from binary evolution models.

In addition to the short duration transients there exists a group of X--ray binaries that return to quiescence after a period of activity that lasted several years to more than a decade (\cite{2001ApJ...560L.159W}; for the list of currently known systems see \cite{2007MNRAS.377.1295J}). In those sources the neutron star crust is not in thermal equilibrium with the core. The pycnonuclear reactions have heated the crust to temperatures higher than those related to thermal equilibrium (\cite{2002ApJ...580..413R}).

Here, I briefly discuss these different ways in which current and future observations of quiescent low--mass X--ray binaries can provide constraints on the neutron star equation of state.

\section{Observed quiescent properties}

In Figure 1 I provide an update to the figure from \cite{2004MNRAS.354..666J}. For this I have used results that have been published since 2004 and the results from the analysis of an \xmm\, observation of 4U~1608--552 in quiescence (Altamirano et al.~in prep; see Figure 2). As can be seen from Figure 1, there seems to be a systematic trend in the contribution of the power--law to the total quiescent luminosity in the 0.5--10 keV band. At luminosities above $\approx 10^{34}$ erg s$^{-1}$ the spectrum can be completely described by a power law spectral component. For decreasing luminosities the power law contribution decreases as well. Near luminosities of a few times 10$^{33}$ erg s$^{-1}$ the power law contributes $\approx 10$\%; most of the emission in the 0.5--10 keV band can be described by a neutron star atmosphere or black body model. Also around this luminosity the correlation between the power--law contribution and the luminosity changes to an anti--correlation. Which means that at  luminosities lower than 10$^{33}$ erg s$^{-1}$ the power law becomes progressively more important. Interestingly, this anti-correlation can be explained if the power--law contribution remains constant and the soft thermal contribution decreases gradually when the luminosity changes from 10$^{33}$ erg s$^{-1}$ to 10$^{32}$ erg s$^{-1}$.

\begin{figure*}
\includegraphics[width=110mm,height=110mm,angle=-90]{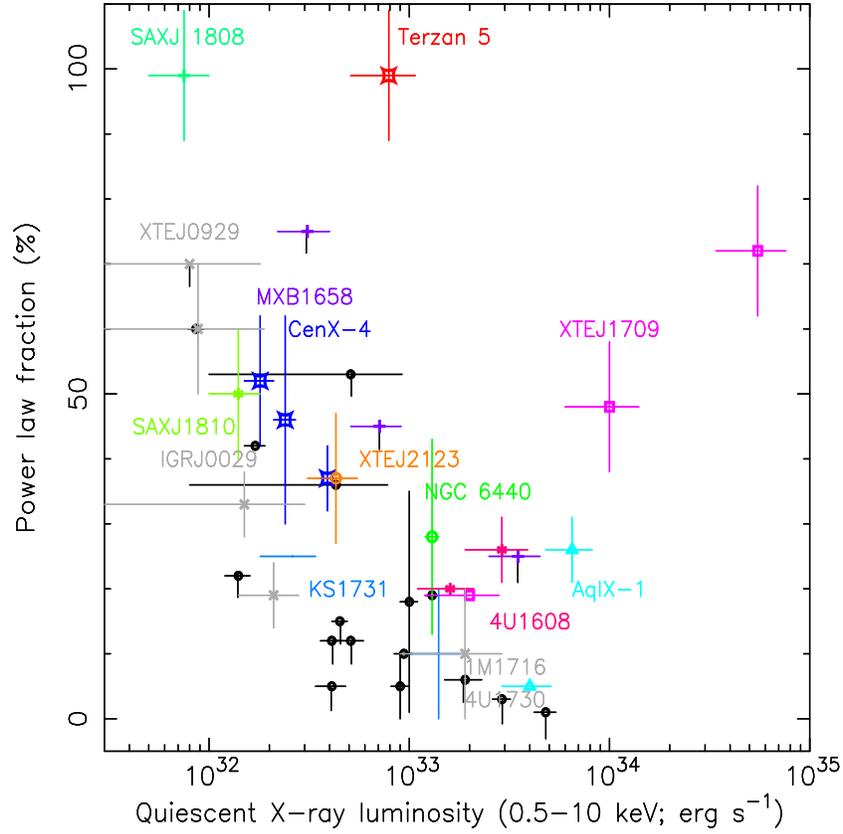}
\caption{An update of the Figure presented in \cite{2004MNRAS.354..666J}, showing the fractional contribution of the power--law spectral component to the 0.5--10 keV quiescent luminosity in the sample of sources bright enough for spectral fitting. The black dots are sources in Globular Clusters (from \cite{2003ApJ...588..452H}). The data points without an upper error bar denote upper limits. Two points can be noticed. First, the quiescent luminosity of some sources is variable. Second, there seems to be a systematic trend in the contribution of the power--law to the total quiescent luminosity; at high quiescent luminosities the spectrum is completely described by a power law spectral component. Near luminosities of a few times 10$^{33}$ erg s$^{-1}$ the power law contributes $\approxlt 10$\%; most of the emission in the 0.5--10 keV band close to this luminosity can be described by a neutron star atmosphere or black body model. At lower luminosities still the power law becomes progressively more important. }
\label{fig1}
\end{figure*}

\begin{figure}
\includegraphics[width=70mm,height=70mm,angle=-90]{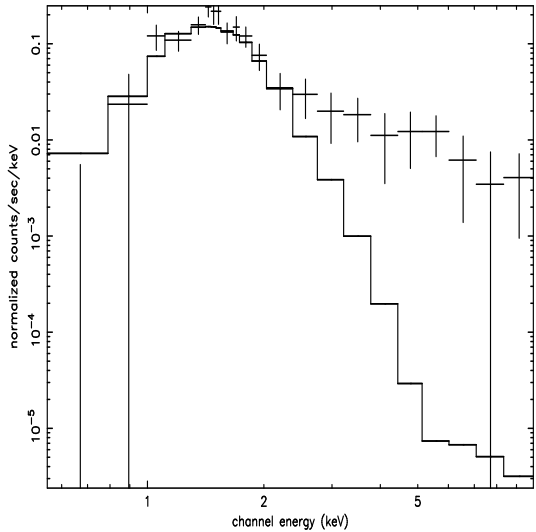}
\caption{The \xmm\, pn spectrum of 4U~1608--552 in quiescence fitted with a fit--function consisting of a neutron star atmosphere model plus a power--law component both subject to interstellar extinction. To show the contribution of the power--law component the normalisation of the power law has been set to zero for the plot. A more detailed description of the results of our analysis of the \xmm\, data of this source will appear in Altamirano et al.~(in preparation).}
\label{fig2}
\end{figure}

\section{Radius measurements of neutron stars?}

Heinke et al.~\cite{2003ApJ...588..452H} and \cite{2007arXiv0708.3816W} find values for the neutron star mass and radius consistent with those of canonical neutron stars using neutron star atmosphere model fits to X--ray spectra of quiescent X--ray binaries in Globular Clusters. However, it is still not certain whether the soft thermal spectral component is really due to the cooling neutron star. Substantial variations on timescales that vary from hundreds of seconds to days and months have been observed in the quiescent X--ray luminosity of 4U~2129+47, Aql~X--1, IGR~J00291+5934, the neutron star transient in quiescence in NGC~6440 (CX1), SAX~J1808.4--3658 and \src\, (cf.~\cite{2000ApJ...529..985R}, \cite{2002ApJ...577..346R}, \cite{2005MNRAS.361..511J}, \cite{2005ApJ...620..922C}, \cite{2007ApJ...660.1424H}, \cite{1997A&A...324..941C}). In the cases where the statistics is best i.e.~Aql~X--1 and Cen~X--4, it is clear that the variability can not be described by only variations in the power law (\cite{2002ApJ...577..346R}, \cite{2004ApJ...601..474C}).

Even though the neutron star mass, radius, distance and temperature cannot vary on short timescales, small observed changes in the neutron star effective temperature can be explained in light of an NSA model if an outburst or a type I X--ray burst took place between the observations that provide the evidence for variability. Small temperature changes can be caused by changes in the heat blanketing layer below the neutron star atmosphere (\cite{2002ApJ...574..920B}). The heat blanketing layer consists of ashes of nuclear burning produced in type I X--ray bursts and of a layer of H and He that remains after an outburst. The thickness of the latter layer varies from outburst to outburst. A thicker layer means a higher heat conductivity which implies a higher observed effective temperature for a given (unchanged) core temperature. However, the luminosity in the soft band in Cen~X--4 varies on time scales too short to be explained by variations in the thickness of the H/He layer.

Currently, the favoured explanation is that the power--law spectral component varies in accord with ${\rm N_H}$ (\cite{2003ApJ...597..474C}, \cite{2004ApJ...601..474C}). In this way the thermal spectral component can be kept constant. If the soft thermal component varies on short time
scales there is a problem with the interpretation that it arises in the NSA. 

\subsection{Usage of spectra obtained with the \xmm\, Reflecting Grating Spectrometer}

Coupled variations between the power--law index and ${\rm N_H}$ such as found in Aql~X--1 by \cite{2003ApJ...597..474C} and in \src\, by \cite{2004ApJ...601..474C} can be accommodated in the model that explains the power law component as shocked emission from the pulsar wind and the infalling material (\cite{1998A&ARv...8..279C}, \cite{2003ApJ...597..474C}) if the amount of (ionised) matter at the shock region and ${\rm N_H}$ are correlated. The nature of this correlation is unclear. Hence, the coupled variability between ${\rm N_H}$ and the power law is rather ad hoc.

With a long \xmm\, observation of \src\, one can determine ${\rm N_H}$ by measuring the equivalent width of the O~$I$ K--edge observable in the RGS spectrum (see Figure 3; figure from \cite{2007xmmconf}). In this way ${\rm N_H}$ can be determined independently from the broadband spectral fit to the EPIC pn and MOS spectra, leaving only the temperature and normalisation of the soft thermal component and the power--law index and normalisation as free parameters to explain the variability.

\begin{figure*}
\includegraphics[width=90mm,height=140mm,angle=-90]{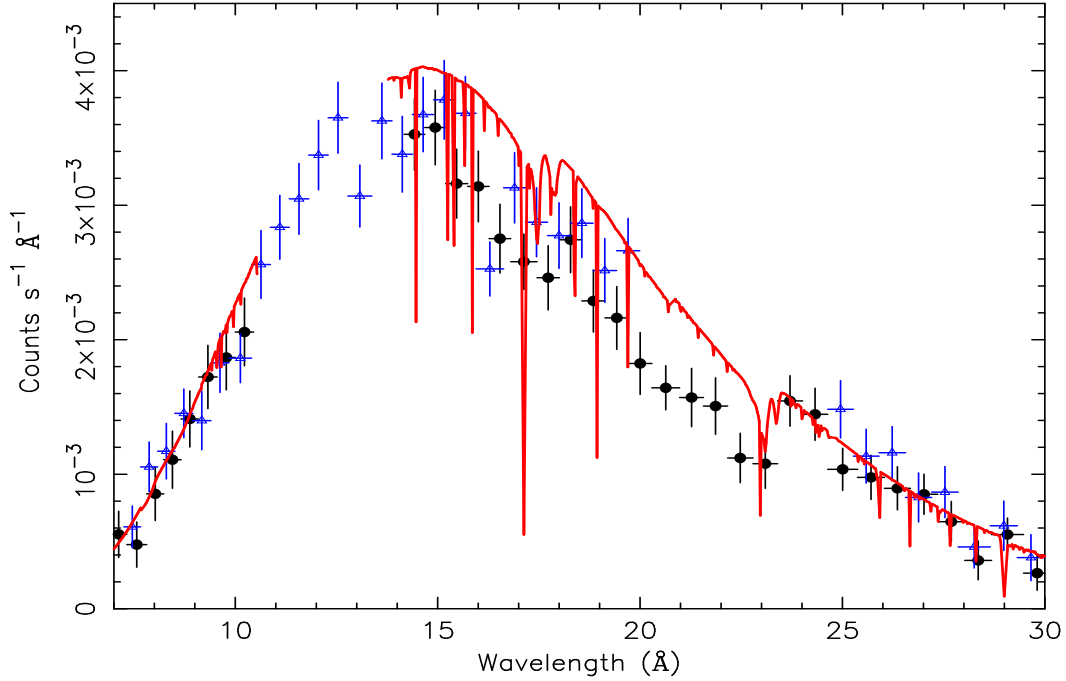}
\caption{A simulated RGS spectrum for a 250 ks--long \xmm\,
observation of \src (from \cite{2007xmmconf}). The black bullets with error bars are the rebinned simulated
RGS--1 data points, the (blue) open triangles with error bars are the rebinned
simulated RGS--2 data points, and the (red), drawn line is the model fitted to the
RGS--1 data where the O column density is put to zero showing the large difference
between the data and the model especially between 15--22\AA. Most of the narrow
lines are due to CCD gaps and detector drop outs. Most of the O~I
K--edge wavelength is not covered by the RGS--2 due to loss of CCD4 in RGS--2. The
O column and, under the assumption of solar abundances, the ${\rm N_H}$ can
be determined with an accuracy of 8--9\% if the broad band spectral components are
held fixed. Such an observation will 
allow one to determine if indeed the factor of 2--3 variation in  ${\rm N_H}$
necessary to explain the luminosity variations as coupled variations in the power
law index and ${\rm N_H}$ (\cite{2004ApJ...601..474C}) is present or not.}
\label{fig3}
\end{figure*}

There are two possible outcomes of such a study. First, if it is determined that indeed
the soft thermal spectral component does not vary and the variability can be
explained by coupled variations in the power law and ${\rm N_H}$ it would boost the
confidence in the masses and radii that have been and can be determined from fitting
the NSA models to the X-ray spectra of quiescent low--mass X--ray binaries. Note
that unknown uncertainties in the neutron star atmosphere models remain. However, in
this respect it is interesting to note that various atmosphere models give similar
results (see discussion in \cite{2007arXiv0708.3816W}).

The second possible outcome is that the soft thermal component varies
substantially. This would either mean that there exists a currently unidentified
mechanism associated with crustal heating that causes the effective temperature to
change on short timescales or that the soft X-ray spectral component is caused by
another process such as residual accretion as considered by \cite{1987A&A...182...47V} 
and \cite{1995ApJ...439..849Z}. 

\section{Undetected quiescent neutron star X--ray transients} 

If the soft thermal component that is observed in many quiescent X--ray transients is {\it not} due to the thermal glow of the hot neutron star core moderated by the NSA this would imply that the cores of these neutron stars are so cold that the hot thermal glow is not detectable, providing evidence for enhanced cooling mechanisms in these neutron stars (cf.~\cite{2007ApJ...665L.147J}for the case of 1H~1905+000). We have identified several sources which have not been detected in quiescence in X--rays (see Table 1). Note that the error on some of the luminosities mentioned in Table 1 can be of the order of a factor 4 due to uncertainties in the distance. 

\begin{table}
\begin{tabular}{lcc}
\hline
  \tablehead{1}{l}{b}{Undetected Sources}
&  \tablehead{1}{r}{b}{Luminosity limit (erg s$^{-1}$)}
&  \tablehead{1}{r}{b}{Reference}\\
\hline
XMMU~J174716.1--281048 & $<$6.2$\times 10^{32}$ & \cite{2003ATel..147....1S}  \\
2S~1803--245 & $<$2.8$\times 10^{32}$ &  \cite{2007MNRAS.380.1637C} \\
EXO~1747--214 & $<$7$\times 10^{31}$ &  \cite{2005ApJ...635.1233T}\\
Terzan 1 & $<$1$\times 10^{33}$ \tablenote{but see \cite{2006MNRAS.369..407C} }& \cite{2002ApJ...572.1002W}\\
XTE~J1751--305 & $<$0.2--2$\times 10^{32}$ &  \cite{2005ApJ...619..492W} \\
XTE~J1807--294 & $<$4$\times 10^{31}$ & \cite{2005AA...434L...9C} \\
1H~1905+000 & $<$2$\times 10^{30}$ & \cite{2006MNRAS.368.1803J}, \cite{2007ApJ...665L.147J} \\
\hline
\end{tabular}
\caption{Neutron star X--ray transients that have not been detected by \cxo\, or \xmm\, while in quiescence.}
\label{tab:a}
\end{table}

In the case of 1H~1905+000 the distance is known with an accuracy of $\approx 20$\% as a radius expansion type~I X--ray burst has been discovered from this source (\cite{1990A&A...228..115C}). The extremely deep limit on the quiescent luminosity implies that the neutron star core probably has access to the direct URCA neutrino emission mechanism. This conclusion can be avoided if neutron star core is not in a steady state. For instance if 
the recurrence time of this transient is of the order of ten thousand years (Brown, E.F. private communication). A neutron star transient that has been detected in quiescence but for which the soft thermal spectral component remains so far undetectable is SAX~J1808.4--3658. A detailed discussion of that source is given by Heinke et al.~in this volume.

\section{Long--duration neutron star X--ray transients in quiescence} 

There exists a group of transients for which the outburst(s) last so long that after an outburst ends the crust is no longer in thermal equilibrium with the core. When the accretion rate decreases/stops the cooling neutron star crust can be observed (e.g.~\cite{2001ApJ...560L.159W}, \cite{2002ApJ...580..413R}, \cite{2006MNRAS.372..479C}). The two sources that have been observed in detail during this cooling phase are KS~1731--260 and MXB~1658--29. The latest results on these sources have been presented by \cite{2006MNRAS.372..479C}. Note that even though these sources are also plotted in Figure 1, their spectra can be completely described by an NSA: only upper limits on a power law contribution have been derived. Furthermore, even if the soft thermal component in Cen~X--4 turns out to be variable this does not necessarily imply that the thermal component observed in these long--duration transients is of the same origin as that in the other (short--duration) transients. Therefore, presently, NSA fits to the spectra of such long--duration transients just after an outburst seem an excellent way to determining the neutron star mass and radius. Unfortunately, KS~1731--260 and MXB~1658--29 suffer from a rather high ${\rm N_H}$ and their distance is large limiting the source flux in the soft X--ray bands. Furthermore, even though their distances have been derived via radius expansion bursts, the distance uncertainty is still too large to provide stringent constraints on the neutron star radius in these two sources. A neutron star low--mass X--ray binary in a nearby globular cluster that has been accreting persistently for several years turning to quiescence would provide an ideal source to follow with \cxo\, and or \xmm.

Cackett et al.~\cite{2006MNRAS.372..479C} present evidence for direct URCA neutrino emission processes taking place in the cores of KS~1731--260 and MXB~1658--29 and for a high termal conductivity of their crusts. Recent results of \cite{2007shternin}, however, show that the observations of KS~1731--260 can be explained with modified URCA and a "normal" crustal thermal conductivity (MXB~1658--29 has not been modelled in detail by these authors yet).


\section{Conclusions}

There are several potential ways in which (X--ray) observations of quiescent low--mass X--ray binaries can provide constraints on the neutron star equation of state via mass and radius determinations, but also via the derived cooling properties. \cxo\, and \xmm\, observations of a future nearby transient that had an outburst long enough to heat the crust significantly out of thermal equilibrium with the core should provide constraints on the neutron star mass and radius if the source distance can be determined accurately. For the more common transients, i.e.~those who returned to quiescence after a weeks to months--long outburst, the observed spectral variability should be investigate further before firm conclusions about the neutron star radius and mass can be drawn from fitting neutron star atmosphere models to the quiescent X--ray spectra. A long \xmm\, observation of e.g.~\src\, would allow the RGS spectrum to be extracted and to investigate if indeed the soft thermal spectral component is stationary. If so, and if the distance to the system is accurately known the neutron star mass and radius can be determined from a NSA model fit to the quiescent spectra.
Finally, the low quiescent (thermal) luminosities of several sources (most notably 1H~1905+000 and SAX~J1808.4--3658) provide evidence for the presence of neutrino emissivities larger than can be provided by the modified URCA process. Presently, it is thought that only neutron stars more massive than the canonical 1.4 \msun\, might have access to these direct URCA processes.


\begin{theacknowledgments}
  PGJ acknowledges support from the Netherlands
Organisation for Scientific Research and insightfull discussions with Ed Brown.
\end{theacknowledgments}

\bibliographystyle{aipproc}

\begin{thebibliography}{47}
\expandafter\ifx\csname natexlab\endcsname\relax\def\natexlab#1{#1}\fi
\providecommand{\enquote}[1]{``#1''}
\expandafter\ifx\csname url\endcsname\relax
  \def\url#1{\texttt{#1}}\fi
\expandafter\ifx\csname urlprefix\endcsname\relax\def\urlprefix{URL }\fi
\providecommand{\eprint}[2][]{\url{#2}}

\bibitem[{Chen} et~al.(1997)]{1997ApJ...491..312C}
W.~{Chen}, C.~R. {Shrader}, and M.~{Livio}, \emph{\apj} \textbf{491}, 312--+
  (1997).

\bibitem[{van Paradijs} et~al.(1987)]{1987A&A...182...47V}
J.~{van Paradijs}, F.~{Verbunt}, R.~A. {Shafer}, and K.~A. {Arnaud},
  \emph{\aap} \textbf{182}, 47--50 (1987).

\bibitem[{McClintock} et~al.(1995)]{1995ApJ...442..358M}
J.~E. {McClintock}, K.~{Horne}, and R.~A. {Remillard}, \emph{\apj}
  \textbf{442}, 358--365 (1995).

\bibitem[{Campana} et~al.(1997)]{1997A&A...324..941C}
S.~{Campana}, S.~{Mereghetti}, L.~{Stella}, and M.~{Colpi}, \emph{\aap}
  \textbf{324}, 941--942 (1997), \eprint{arXiv:astro-ph/9701234}.

\bibitem[{Wijnands} et~al.(2001)]{2001ApJ...560L.159W}
R.~{Wijnands}, J.~M. {Miller}, C.~{Markwardt}, W.~H.~G. {Lewin}, and M.~{van
  der Klis}, \emph{\apjl} \textbf{560}, L159--L162 (2001).

\bibitem[{Rutledge} et~al.(2002{\natexlab{a}})]{2002ApJ...580..413R}
R.~E. {Rutledge}, L.~{Bildsten}, E.~F. {Brown}, G.~G. {Pavlov}, V.~E. {Zavlin},
  and G.~{Ushomirsky}, \emph{\apj} \textbf{580}, 413--422 (2002{\natexlab{a}}).

\bibitem[{Campana} et~al.(2002)]{2002ApJ...575L..15C}
S.~{Campana}, L.~{Stella}, F.~{Gastaldello}, S.~{Mereghetti}, M.~{Colpi}, G.~L.
  {Israel}, L.~{Burderi}, T.~{Di Salvo}, and R.~N. {Robba}, \emph{\apjl}
  \textbf{575}, L15--L19 (2002),
  \urlprefix\url{http://cdsads.u-strasbg.fr/cgi-bin/nph-bib_query?bibcode=2002%
ApJ...575L..15C&db_key=AST}.

\bibitem[{Jonker} et~al.(2003)]{2003MNRAS.341..823J}
P.~G. {Jonker}, M.~{M{\' e}ndez}, G.~{Nelemans}, R.~{Wijnands}, and M.~{van der
  Klis}, \emph{\mnras} \textbf{341}, 823--831 (2003).

\bibitem[{Tomsick} et~al.(2004)]{2004ApJ...610..933T}
J.~A. {Tomsick}, D.~M. {Gelino}, J.~P. {Halpern}, and P.~{Kaaret}, \emph{\apj}
  \textbf{610}, 933--940 (2004).

\bibitem[{Campana} et~al.(2004)]{2004ApJ...601..474C}
S.~{Campana}, G.~L. {Israel}, L.~{Stella}, F.~{Gastaldello}, and
  S.~{Mereghetti}, \emph{\apj} \textbf{601}, 474--478 (2004).

\bibitem[{Rutledge} et~al.(2002{\natexlab{b}})]{2002ApJ...577..346R}
R.~E. {Rutledge}, L.~{Bildsten}, E.~F. {Brown}, G.~G. {Pavlov}, and V.~E.
  {Zavlin}, \emph{\apj} \textbf{577}, 346--358 (2002{\natexlab{b}}).

\bibitem[{Brown} et~al.(1998)]{1998ApJ...504L..95B}
E.~F. {Brown}, L.~{Bildsten}, and R.~E. {Rutledge}, \emph{\apjl} \textbf{504},
  L95--+ (1998),
  \urlprefix\url{http://adsabs.harvard.edu/cgi-bin/nph-bib_query?bibcode=1998A%
pJ...504L..95B&db_key=AST}.

\bibitem[{Colpi} et~al.(2001)]{2001ApJ...548L.175C}
M.~{Colpi}, U.~{Geppert}, D.~{Page}, and A.~{Possenti}, \emph{\apjl}
  \textbf{548}, L175--L178 (2001),
  \urlprefix\url{http://adsabs.harvard.edu/cgi-bin/nph-bib_query?bibcode=2001A%
pJ...548L.175C&db_key=AST}.

\bibitem[{Haensel} and {Zdunik}(1990)]{1990A&A...227..431H}
P.~{Haensel}, and J.~L. {Zdunik}, \emph{\aap} \textbf{227}, 431--436 (1990).

\bibitem[{Kraft} et~al.(1962)]{1962ApJ...136..312K}
R.~P. {Kraft}, J.~{Mathews}, and J.~L. {Greenstein}, \emph{\apj} \textbf{136},
  312--+ (1962).

\bibitem[{Salpeter} and {van Horn}(1969)]{1969ApJ...155..183S}
E.~E. {Salpeter}, and H.~M. {van Horn}, \emph{\apj} \textbf{155}, 183--+
  (1969).

\bibitem[{Kitamura}(2000)]{2000ApJ...539..888K}
H.~{Kitamura}, \emph{\apj} \textbf{539}, 888--901 (2000).

\bibitem[{Yakovlev} and {Pethick}(2004)]{2004ARA&A..42..169Y}
D.~G. {Yakovlev}, and C.~J. {Pethick}, \emph{\araa} \textbf{42}, 169--210
  (2004).

\bibitem[{Yakovlev}(2007)]{yakovthisvol}
D.~G. {Yakovlev}, \emph{this volume}  (2007).

\bibitem[{Zavlin} et~al.(1996)]{1996A&A...315..141Z}
V.~E. {Zavlin}, G.~G. {Pavlov}, and Y.~A. {Shibanov}, \emph{\aap} \textbf{315},
  141--152 (1996),
  \urlprefix\url{http://adsabs.harvard.edu/cgi-bin/nph-bib_query?bibcode=1996%
A%26A...315..141Z&db_key=AST}.

\bibitem[{G{\" a}nsicke} et~al.(2002)]{2002A&A...386.1001G}
B.~T. {G{\" a}nsicke}, T.~M. {Braje}, and R.~W. {Romani}, \emph{\aap}
  \textbf{386}, 1001--1008 (2002).

\bibitem[{Heinke} et~al.(2006)]{2006ApJ...644.1090H}
C.~O. {Heinke}, G.~B. {Rybicki}, R.~{Narayan}, and J.~E. {Grindlay},
  \emph{\apj} \textbf{644}, 1090--1103 (2006), \eprint{arXiv:astro-ph/0506563}.

\bibitem[{Heinke} et~al.(2003)]{2003ApJ...588..452H}
C.~O. {Heinke}, J.~E. {Grindlay}, D.~A. {Lloyd}, and P.~D. {Edmonds},
  \emph{\apj} \textbf{588}, 452--463 (2003).

\bibitem[{Webb} and {Barret}(2007)]{2007arXiv0708.3816W}
N.~A. {Webb}, and D.~{Barret}, \emph{ArXiv e-prints} \textbf{708} (2007),
  \eprint{0708.3816}.

\bibitem[{Heinke} et~al.(2007)]{2007ApJ...660.1424H}
C.~O. {Heinke}, P.~G. {Jonker}, R.~{Wijnands}, and R.~E. {Taam}, \emph{\apj}
  \textbf{660}, 1424--1427 (2007), \eprint{arXiv:astro-ph/0612232}.

\bibitem[{Jonker} et~al.(2006)]{2006MNRAS.368.1803J}
P.~G. {Jonker}, C.~G. {Bassa}, G.~{Nelemans}, A.~M. {Juett}, E.~F. {Brown}, and
  D.~{Chakrabarty}, \emph{\mnras} \textbf{368}, 1803--1810 (2006),
  \eprint{astro-ph/0602625}.

\bibitem[{Jonker} et~al.(2007{\natexlab{a}})]{2007ApJ...665L.147J}
P.~G. {Jonker}, D.~{Steeghs}, D.~{Chakrabarty}, and A.~M. {Juett}, \emph{\apjl}
  \textbf{665}, L147--L150 (2007{\natexlab{a}}), \eprint{arXiv:0706.3421}.

\bibitem[{Jonker} et~al.(2007{\natexlab{b}})]{2007MNRAS.377.1295J}
P.~G. {Jonker}, C.~G. {Bassa}, and S.~{Wachter}, \emph{\mnras} \textbf{377},
  1295--1300 (2007{\natexlab{b}}), \eprint{arXiv:astro-ph/0703020}.

\bibitem[{Jonker} et~al.(2004)]{2004MNRAS.354..666J}
P.~G. {Jonker}, D.~K. {Galloway}, J.~E. {McClintock}, M.~{Buxton}, M.~{Garcia},
  and S.~{Murray}, \emph{\mnras} \textbf{354}, 666--674 (2004).

\bibitem[{Rutledge} et~al.(2000)]{2000ApJ...529..985R}
R.~E. {Rutledge}, L.~{Bildsten}, E.~F. {Brown}, G.~G. {Pavlov}, and V.~E.
  {Zavlin}, \emph{\apj} \textbf{529}, 985--996 (2000),
  \eprint{arXiv:astro-ph/9909319}.

\bibitem[{Jonker} et~al.(2005)]{2005MNRAS.361..511J}
P.~G. {Jonker}, S.~{Campana}, D.~{Steeghs}, M.~A.~P. {Torres}, D.~K.
  {Galloway}, C.~B. {Markwardt}, D.~{Chakrabarty}, and J.~{Swank},
  \emph{\mnras} \textbf{361}, 511--516 (2005), \eprint{arXiv:astro-ph/0505120}.

\bibitem[{Cackett} et~al.(2005)]{2005ApJ...620..922C}
E.~M. {Cackett}, R.~{Wijnands}, C.~O. {Heinke}, P.~D. {Edmonds}, W.~H.~G.
  {Lewin}, D.~{Pooley}, J.~E. {Grindlay}, P.~G. {Jonker}, and J.~M. {Miller},
  \emph{\apj} \textbf{620}, 922--928 (2005).

\bibitem[{Brown} et~al.(2002)]{2002ApJ...574..920B}
E.~F. {Brown}, L.~{Bildsten}, and P.~{Chang}, \emph{\apj} \textbf{574},
  920--929 (2002).

\bibitem[{Campana} and {Stella}(2003)]{2003ApJ...597..474C}
S.~{Campana}, and L.~{Stella}, \emph{\apj} \textbf{597}, 474--478 (2003).

\bibitem[{Campana} et~al.(1998)]{1998A&ARv...8..279C}
S.~{Campana}, M.~{Colpi}, S.~{Mereghetti}, L.~{Stella}, and M.~{Tavani},
  \emph{\araa} \textbf{8}, 279--316 (1998),
  \urlprefix\url{http://adsabs.harvard.edu/cgi-bin/nph-bib_query?bibcode=1998%
A%26ARv...8..279C&db_key=AST}.

\bibitem[{Jonker} et~al.(2007{\natexlab{c}})]{2007xmmconf}
P.~G. {Jonker}, J.~{Kaastra}, M.~{M\'{e}ndez}, and J.~J.~M. {In 't Zand},
  \emph{Astronomische Nachrichten}  (2007{\natexlab{c}}).

\bibitem[{Zampieri} et~al.(1995)]{1995ApJ...439..849Z}
L.~{Zampieri}, R.~{Turolla}, S.~{Zane}, and A.~{Treves}, \emph{\apj}
  \textbf{439}, 849--853 (1995),
  \urlprefix\url{http://adsabs.harvard.edu/cgi-bin/nph-bib_query?bibcode=1995A%
pJ...439..849Z&db_key=AST}.

\bibitem[{Sidoli} and {Mereghetti}(2003)]{2003ATel..147....1S}
L.~{Sidoli}, and S.~{Mereghetti}, \emph{The Astronomer's Telegram}
  \textbf{147}, 1--+ (2003).

\bibitem[{Cornelisse} et~al.(2007)]{2007MNRAS.380.1637C}
R.~{Cornelisse}, R.~{Wijnands}, and J.~{Homan}, \emph{\mnras} \textbf{380},
  1637--1641 (2007), \eprint{arXiv:0707.0995}.

\bibitem[{Tomsick} et~al.(2005)]{2005ApJ...635.1233T}
J.~A. {Tomsick}, D.~M. {Gelino}, and P.~{Kaaret}, \emph{\apj} \textbf{635},
  1233--1238 (2005).

\bibitem[{Wijnands} et~al.(2002)]{2002ApJ...572.1002W}
R.~{Wijnands}, C.~O. {Heinke}, and J.~E. {Grindlay}, \emph{\apj} \textbf{572},
  1002--1005 (2002), \eprint{arXiv:astro-ph/0111337}.

\bibitem[{Wijnands} et~al.(2005)]{2005ApJ...619..492W}
R.~{Wijnands}, J.~{Homan}, C.~O. {Heinke}, J.~M. {Miller}, and W.~H.~G.
  {Lewin}, \emph{\apj} \textbf{619}, 492--502 (2005).

\bibitem[{Campana} et~al.(2005)]{2005AA...434L...9C}
S.~{Campana}, N.~{Ferrari}, L.~{Stella}, and G.~L. {Israel}, \emph{\aap}
  \textbf{434}, L9--L12 (2005).

\bibitem[{Cackett} et~al.(2006{\natexlab{a}})]{2006MNRAS.369..407C}
E.~M. {Cackett}, R.~{Wijnands}, C.~O. {Heinke}, D.~{Pooley}, W.~H.~G. {Lewin},
  J.~E. {Grindlay}, P.~D. {Edmonds}, P.~G. {Jonker}, and J.~M. {Miller},
  \emph{\mnras} \textbf{369}, 407--415 (2006{\natexlab{a}}),
  \eprint{arXiv:astro-ph/0512168}.

\bibitem[{Chevalier} and {Ilovaisky}(1990)]{1990A&A...228..115C}
C.~{Chevalier}, and S.~A. {Ilovaisky}, \emph{\aap} \textbf{228}, 115--124
  (1990).

\bibitem[{Cackett} et~al.(2006{\natexlab{b}})]{2006MNRAS.372..479C}
E.~M. {Cackett}, R.~{Wijnands}, M.~{Linares}, J.~M. {Miller}, J.~{Homan}, and
  W.~H.~G. {Lewin}, \emph{\mnras} \textbf{372}, 479--488 (2006{\natexlab{b}}),
  \eprint{arXiv:astro-ph/0605490}.

\bibitem[{Shternin} et~al.(2007)]{2007shternin}
P.~S. {Shternin}, D.~G. {Yakovlev}, P.~{Haensel}, and A.~Y. {Potekhin},
  \emph{ArXiv e-prints} \textbf{86} (2007), \eprint{0708.0086}.

\end{thebibliography}


\end{document}